\documentclass[12pt]{article}

\def\gtwid{\mathrel{\raise.3ex\hbox{$>$\kern-.75em\lower1ex\hbox{$\sim$}}}}
\def\ltwid{\mathrel{\raise.3ex\hbox{$<$\kern-.75em\lower1ex\hbox{$\sim$}}}}
\def\square{\kern1pt\vbox{\hrule height 1.2pt\hbox{\vrule width 1.2pt\hskip 3pt
   \vbox{\vskip 6pt}\hskip 3pt\vrule width 0.6pt}\hrule height 0.6pt}\kern1pt}

\begin{document}

\begin{titlepage}

\begin{flushright}
UFIFT-QG-15-04
\end{flushright}

\vskip 2cm

\begin{center}
{\bf Some Inconvenient Truths}
\end{center}

\vskip 2cm

\begin{center}
R. P. Woodard$^{\dagger}$
\end{center}

\begin{center}
\it{Department of Physics, University of Florida, Gainesville, FL 32611}
\end{center}

\vspace{1cm}

\begin{center}
ABSTRACT
\end{center}
A recent paper by Fr\"ob employs the linearized Weyl-Weyl correlator
to construct the tensor power spectrum. Although his purpose was to argue 
that infrared divergences and secular growth in the graviton propagator 
are gauge artefacts, a closer examination of the problem leads to the
opposite conclusion. The analogies with the BMS symmetries of graviton 
scattering on a flat background, and with the Aharonov-Bohm effect of 
quantum mechanics, suggest that de Sitter breaking secular growth is 
likely to be observable in graviton loop effects. And a recent result 
for the vacuum polarization does seem to show it.

\begin{flushleft}
\end{flushleft}

\vskip 2cm

\begin{flushleft}
$^{\dagger}$ e-mail: woodard@phys.ufl.edu
\end{flushleft}

\end{titlepage}

\section{Introduction}

There has been a long controversy about de Sitter breaking by the 
graviton propagator on de Sitter background
\cite{Antoniadis:1986sb,Allen:1986tt,Tsamis:1992xa,Kleppe:1993fz,
Hawking:2000ee,Higuchi:2001uv,Kouris:2001hz,Higuchi:2002sc,
Woodard:2004ut,Higuchi:2008fu,Faizal:2008ns,Marolf:2008hg,Miao:2009hb,
Miao:2010vs,Miao:2011fc,Faizal:2011iv,Higuchi:2011vw,Miao:2011ng,
Kahya:2011sy,Mora:2012kr,Higuchi:2012vy,Mora:2012zh,Mora:2012zi,
Morrison:2013rqa,Miao:2013isa,Bernar:2014lna}. Much has been learned 
that is valid no matter which view prevails. Significant insights 
include:
\begin{itemize}
\item{A linearization instability precludes adding de Sitter invariant 
gauge fixing terms to the action, although enforcing de Sitter invariant 
exact gauge conditions as strong operator equations is allowed 
\cite{Miao:2009hb}.}
\item{The best way of expressing the tensor structure of a graviton
propagator in an exact gauge is as a sum of differential projectors 
acting on scalar structure functions which obey simple equations
\cite{Miao:2011fc,Mora:2012zi}.}
\item{The entire difference between the de Sitter invariant and de
Sitter breaking constructions in the same fully-fixed gauge can be 
traced to what these scalar equations are and how one solves them. 
Further, this difference takes the form of a gauge transformation 
\cite{Morrison:2013rqa}.}
\item{A solution to the propagator equation need not correspond to
a propagator in the sense of being the expectation value of the
time-ordered product of two fields in the presence of some positive
norm state \cite{Tsamis:2000ah,Janssen:2008px}.}
\item{Analytic continuation fails to register power law infrared 
divergences \cite{Folacci:1992xc,Janssen:2008px,Miao:2010vs} and 
can convert positive into negative norm states \cite{Miao:2013isa}.}
\end{itemize}
 
A particularly important insight has been the close relation between
the tensor power spectrum of primordial inflation and the graviton 
propagator in transverse-traceless-synchronous (TTS) gauge, which poses 
a formidable problem for those who deny de Sitter breaking
\cite{Miao:2011ng,Miao:2013isa}. A recent paper by Fr\"ob confronts
this issue by using spatial Fourier transforms of the linearized 
Weyl-Weyl correlator to reconstruct the tensor power spectrum 
\cite{Frob:2014fqa}. Because the linearized Weyl-Weyl correlator for de 
Sitter is infrared finite and free of secular dependence
\cite{Kouris:2001hz,Mora:2012kr,Mora:2012zh}, Fr\"ob argues that both of 
these features in the TTS gauge propagator are gauge artefacts. He 
further argues that one can consistently construct the tensor power 
spectrum using the infrared finite propagator obtained by Higuchi, Marolf 
and Morrison by subjecting the TTS gauge result to a coordinate 
transformation which grows at spatial infinity but preserves the TTS 
condition \cite{Higuchi:2011vw}. 

Fr\"ob's attention to this problem is praise-worthy. However, a fuller 
examination of the facts leads to different conclusions, in particular:
\begin{itemize}
\item{It is not invariance but derivatives which moderate infrared
divergences;}
\item{Infrared divergences are distinct from secular dependence;}
\item{If the validity of spatial Fourier transforms is assumed then
the linearized Weyl-Weyl correlator implies that the usual result for 
the graviton propagator is unique in TTS gauge;}
\item{Conversely, if the validity of large coordinate transformations
is asserted then the linearized Weyl-Weyl correlator does not provide
a unique result for the tensor power spectrum;}
\item{The fact that the symmetries of Bondi, van der Burg and Metzner
\cite{Bondi:1962px} and Sachs \cite{Sachs:1962wk} (hencforth BMS) in 
graviton scattering on flat background act nontrivially on S-matrix 
elements \cite{Strominger:2014pwa,Strominger:2013jfa,He:2014laa} seems 
to imply that we should not consider diffeomorphisms whose parameters 
grow at spatial infinity to be gauge transformations; and}
\item{The analogy with the Aharonov-Bohm effect of quantum mechanics
\cite{Aharonov:1959fk} suggests that secular dependence in the graviton 
propagator can have physical consequences, even though it takes the 
form of a gauge transformation of a completely gauge-fixed quantity.}
\end{itemize}
Section 2 of this paper reviews the TTS gauge propagator and its
relation to the graviton power spectrum. Section 3 gives the arguments 
supporting the conclusions stated above. Section 4 summarizes the 
progress which has been made in the debate about de Sitter breaking.

\section{The Power Spectrum and TTS Gauge}

The tensor power spectrum is extracted by taking the late time limit of 
the temporally coincident graviton 2-point function,
\begin{equation}
\Delta^2_{h}(\eta,k) \equiv \frac{k^3}{2 \pi^2} \int \!\! d^3x e^{-i
\vec{k} \cdot \vec{x}} \Bigl\langle \Omega \Bigl\vert 
h_{ij}(\eta,\vec{x}) h_{ij}(\eta,\vec{0}) \Bigr\vert \Omega 
\Bigr\rangle \; ,
\end{equation}
in the same TTS gauge that was long ago applied to cosmology by Lifshitz 
\cite{Lifshitz:1945du}. One defines the graviton field 
$h_{\mu\nu}(\eta,\vec{x})$ (also the Hubble parameter $H(\eta)$ and the 
first slow roll parameter $\epsilon(\eta)$) in open conformal coordinates,
\begin{equation}
g_{\mu\nu} \equiv a^2 \Bigl[\eta_{\mu\nu} + h_{\mu\nu}\Bigr] \qquad ,
\qquad H(\eta) \equiv \frac{a'(\eta)}{a^2(\eta)} \qquad , \qquad
\epsilon(\eta) \equiv -\frac{H'}{a H^2} \; . \label{geometry}
\end{equation}
De Sitter corresponds to the special case of $\epsilon = 0$ but I will
work with general $\epsilon(\eta)$. One reaches TTS gauge by first 
imposing the volume gauge condition,
\begin{equation}
F_{\mu}(\eta,\vec{x}) \equiv \eta^{\rho\sigma} \Bigl[ h_{\mu\rho ,\sigma}
- \frac12 \partial_{\mu} h_{\rho\sigma} + 2 H a h_{\mu \rho} 
\delta^0_{\sigma}\Bigr] = 0 \; .
\end{equation}
Then the linearized Einstein equation, plus a {\it complete exhaustion of 
all residual gauge freedom in spatial Fourier space} to eliminate $h_{00}$ 
and $h_{0i}$, allows one to express $h_{ij}$ as \cite{Tsamis:1992zt,
Iliopoulos:1998wq},
\begin{eqnarray}
\lefteqn{h_{ij}(\eta,\vec{x}) = \sqrt{32\pi G} \! \int \!\! 
\frac{d^3k}{(2\pi)^3} \sum_{\lambda=\pm 2} \Biggl\{ u(\eta,k) e^{i \vec{k} 
\cdot \vec{x}} \varepsilon_{ij}(\vec{k},\lambda) \alpha(\vec{k},\lambda)}
\nonumber \\
& & \hspace{6cm} + u^*(\eta,k) e^{-i \vec{k} \cdot \vec{x}} 
\varepsilon^*_{ij}(\vec{k},\lambda) \alpha^{\dagger}(\vec{k},\lambda)
\Biggr\} \; . \qquad \label{modesum}
\end{eqnarray}
The transverse-traceless polarization tensors 
$\varepsilon_{ij}(\vec{k},\lambda)$ are identical to those of flat space,
and the mode function $u(\eta,k)$ obeys the same equation as a massless,
minimally coupled (MMC) scalar \cite{Lifshitz:1945du},
\begin{equation}
u'' + 2 H a u' + k^2 u = 0 \qquad , \qquad u {u'}^* - u' u^* = 
\frac{i}{a^2} \; . \label{modeeqn}
\end{equation}
The vacuum state obeys $\alpha(\vec{k},\lambda) \vert \Omega \rangle = 0$
and canonical quantization implies,
\begin{equation}
\Bigl[ \alpha(\vec{k},\lambda) , \alpha^{\dagger}(\vec{k}',\lambda')
\Bigr] = \delta_{\lambda \lambda'} (2\pi)^3 \delta^3( \vec{k} \!-\!
\vec{k}') \; .
\end{equation}

Putting everything together gives a simple result for $\Delta^2_h(\eta,k)$,
\begin{equation}
\Delta^2_{h}(\eta,k) = \frac{k^3}{2 \pi^2} \times 32\pi G \times 2 
\times \Bigl\vert u(\eta,k) \Bigr\vert^2 = \frac{32}{\pi} G k^3 
\Bigl\vert u(\eta,k)\Bigr\vert^2 \; . \label{power}
\end{equation}
The TTS gauge propagator is,
\begin{eqnarray}
i \Bigl[\mbox{}_{ij} \Delta_{k\ell}\Bigr](x;x') & \equiv &
\Bigl\langle \Omega \Bigl\vert T\Bigl[ h_{ij}(\eta,\vec{x}) 
h_{k\ell}(\eta',\vec{x}') \Bigr] \Bigr\vert \Omega \Bigr\rangle \; , \\
& = & 16\pi G \Bigl[ \Pi_{ik} \Pi_{j\ell} \!+\! \Pi_{i\ell} \Pi_{jk}
\!-\! \Pi_{ij} \Pi_{k\ell} \Bigr] i \Delta(x;x') \; , \qquad \label{TTS1}
\end{eqnarray}
where $\Pi_{ij} \equiv \delta_{ij} - \frac{\partial_i \partial_j}{
\nabla^2}$ is the transverse projection operator and $i\Delta(x;x')$
is the propagator for a MMC scalar,
\begin{equation}
i\Delta(x;x') = \!\!\int\!\! \frac{d^3k}{(2\pi)^3} \, e^{i\vec{k} \cdot
(\vec{x} - \vec{x}')} \Bigl[ \theta(\eta \!-\! \eta') u(\eta,k)
u^*(\eta',k) + \theta(\eta' \!-\! \eta) u^*(\eta,k) u(\eta',k) \Bigr] .
\label{TTS2}
\end{equation}
The relation between the power spectrum and the TTS propagator is,
\begin{eqnarray}
\Delta^2_h(\eta,k) & = & \frac{k^3}{2\pi^2} \!\int \!\! d^3x e^{-i\vec{k}
\cdot \vec{x} } \times i\Bigl[\mbox{}_{ij} \Delta_{k\ell}
\Bigr](\eta,\vec{x};\eta,\vec{0}) \; , \\
& = & \frac{32}{\pi} G k^3 \!\! \int \!\! d^3x e^{-i\vec{k} \cdot \vec{x} } 
i\Delta(\eta,\vec{x};\eta,\vec{0}) \; . \label{DtoDh}
\end{eqnarray}
Inverting relation (\ref{DtoDh}) gives the coincident time MMC scalar
propagator, and hence also the TTS gauge graviton propagator,
\begin{equation}
i\Delta(\eta,\vec{x};\eta,\vec{0}) = \int_0^{\infty} \!\! \frac{dk}{k}
\frac{ \Delta^2_h(\eta,k)}{64 \pi G} \; . \label{DhtoD}
\end{equation}

No very explicit solution to (\ref{modeeqn}) is known for general 
$\epsilon(\eta)$ \cite{Tsamis:2002qk} but the constant $\epsilon(\eta)$
solution is,
\begin{equation}
\epsilon' = 0 \quad \Longrightarrow \quad u(\eta,k) = \sqrt{\frac{\pi}{
4 (1 \!-\! \epsilon) H a^3}} \, H^{(1)}_{\nu}\Bigl( \frac{k}{(1 \!-\! 
\epsilon) H a}\Bigr) \quad , \quad \nu \equiv \frac12 \Bigl(\frac{3 \!-\!
\epsilon}{1 \!-\! \epsilon}\Bigr) \; . \label{constepssol}
\end{equation} 
For the inflationary case of $0 \leq \epsilon < 1$ the late time limiting
form is,
\begin{equation}
\epsilon' = 0 \quad \Longrightarrow \quad u(\eta,k) \longrightarrow -
\frac{i \Gamma(\nu)}{\sqrt{4\pi}} \Bigl[ (1\!-\!\epsilon) H a^{\epsilon}
\Bigr]^{\frac1{1-\epsilon}} \Bigl( \frac{2}{k}\Bigr)^{\nu} \; .
\label{constepslim}
\end{equation}
Expression (\ref{constepslim}) is also the limiting form for small $k$.
Ford and Parker used this (and the limiting form for $\epsilon > 1$)
to show that the mode sum for $i\Delta(x;x')$ --- and hence also for the 
TTS gauge propagator --- is infrared divergent for all constant $\epsilon$ 
geometries in the range $0 \leq \epsilon \leq \frac32$ \cite{Ford:1977in}.

Expression (\ref{geometry}) implies $\partial_0 [H a^{\epsilon}] = 
\epsilon' H^2 a^{1+\epsilon}$, so $H a^{\epsilon}$ is constant for constant
$\epsilon$. It is usual to evaluate this constant at the time $\eta_k$ of 
first horizon crossing when $k = H(\eta_k) a(\eta_k)$,
\begin{equation}
\epsilon' = 0 \qquad \Longrightarrow \qquad H(\eta) [a(\eta)]^{\epsilon} =
H(\eta_k) [a(\eta_k)]^{\epsilon} = k^{\epsilon} [H(\eta_k)]^{1-\epsilon} 
\; . \label{constepsrel}
\end{equation}
Substituting relations (\ref{constepslim}) and (\ref{constepsrel}) allows
us to compute the late time limit of the tensor power spectrum for constant
$\epsilon$,
\begin{equation}
\epsilon' = 0 \qquad \Longrightarrow \qquad \Delta^2_h(\eta,k) 
\longrightarrow \frac{16}{\pi} G H^2(\eta_k) \times \frac{4 \Gamma^2(\nu)}{
\pi} \Bigl[2^{\epsilon} (1 \!-\! \epsilon)\Bigr]^{\frac2{1-\epsilon}} \; .
\end{equation}

\section{Facts Are Stubborn Things}

In this section I marshal facts from the previous section to support my
views on infrared divergences and de Sitter breaking secular dependence. 

\subsection{Derivatives moderate the IR, not invariance}

It is not the fact that the linearized Weyl tensor is invariant which renders
its 2-point function infrared finite for de Sitter, but rather the presence
of derivatives. To see this, note from the small $k$ limiting form 
(\ref{constepslim}) that the infrared divergence of $i\Delta(x;x')$ (and 
hence also the TTS gauge propagator) is only logarithmic for the de Sitter
case of $\epsilon = 0$. This means only a {\it single} derivative is needed 
to eliminate it. In particular, there is no infrared divergence in the 
noninvariant quantity,
\begin{equation}
\frac{\partial}{\partial x^m} \, i\Bigl[\mbox{}_{ij} \Delta_{k\ell}
\Bigr](x;x') = \Bigl\langle \Omega \Bigl\vert T\Bigl[ h_{ij , m}(x)
h_{k \ell}(x')\Bigr] \Bigr\vert \Omega \Bigr\rangle \; .
\end{equation}
A single temporal derivative would also produce an infrared finite mode sum 
because the right hand side of expression (\ref{constepslim}) is constant.

It is not even true that invariance guarantees infrared finiteness for 
other constant values of $\epsilon$. From the small $k$ limiting form 
(\ref{constepslim}) one can see that the mode sum for the propagator goes
like,
\begin{equation}
i\Delta(x;x') \sim \int \!\! dk k^2 \times \frac1{k^{2\nu}} = \int \!\! 
\frac{dk}{k} \Bigl( \frac1{k}\Bigr)^{\frac{2\epsilon}{1-\epsilon}} \; . 
\label{smallk}
\end{equation}
The linearized Weyl-Weyl correlator contains four derivatives, so its mode
sum will be infrared divergent for any value of $\epsilon$ in the range
$\frac23 \leq \epsilon \leq \frac54$, with the upper limit derived by 
generalizing (\ref{constepslim}) to the decelerating case of $\epsilon > 
1$.

\subsection{Secular growth is distinct from IR divergences}

Infrared divergences (which only occur in open coordinates) are quite 
distinct from secular growth, which occurs even in closed coordinates
\cite{Higuchi:2002sc}. Secular growth manifests in both coordinate 
systems because higher and higher modes approach the constant 
(\ref{constepslim}) at late times, which is not even different between
the two coordinate systems for high modes \cite{Miao:2013isa}. The 
infrared has nothing at all to do with it, despite the fact that the
late time limiting form (\ref{constepslim}) is the same as the limiting
form for small $k$. Allen and Folacci \cite{Allen:1987tz} demonstrated 
the distinction for the discrete MMC scalar mode sum on closed 
coordinates by expunging the 0-mode and finding precisely the secular 
growth previously obtained in open coordinates by Vilenkin and Ford 
\cite{Vilenkin:1982wt}, Linde \cite{Linde:1982uu} and by Starobinsky 
\cite{Starobinsky:1982ee}.

In contrast, infrared divergences are associated with large numbers of 
small $k$ modes, and with the initial conditions \cite{Fulling:1978ht}. 
At any fixed open coordinates time there are an infinite number of 
super-horizon modes already near the limiting form (\ref{constepslim}), 
whereas the number of super-horizon modes is finite for any fixed closed 
coordinate time \cite{Miao:2013isa}.

Note also that a single time derivative of the graviton propagator
eliminates its IR divergence (on de Sitter) but not its secular growth. 
If one regulates the infrared problem on de Sitter so as to preserve
homogeneity and isotropy then the MMC scalar propagator acquires an
extra term involving $\ln[a(\eta) a(\eta')]$ \cite{Onemli:2002hr}. It 
requires derivatives with respect to both coordinates to annihilate 
this contribution,
\begin{equation}
\partial_{\mu} \ln[a(\eta) a(\eta')] = H a \delta^0_{\mu} \qquad ,
\qquad \partial_{\mu} \partial_{\nu}' \ln[a(\eta) a(\eta')] = 0 \; .
\end{equation}
Even more derivatives are needed to eliminate the extra terms which 
appear for larger values of $\epsilon < 1$ \cite{Iliopoulos:1998wq,
Janssen:2008px}. And differentiated propagators do matter: the 
secular growth experienced at one loop by massless fermions on de 
Sitter derives entirely from diagrams for which the graviton 
propagator is acted upon by a single derivative \cite{Miao:2008sp}.

\subsection{Either Fourier transforms exist or not, pick one}

Either the spatial Fourier transform of $h_{ij}(t,\vec{x})$ exists or
it does not, and neither possibility supports de Sitter invariance. A
different conclusion is only possible by inconsistently using spatial 
Fourier transforms to convert the linearized Weyl-Weyl correlator into 
$\Delta^2_h(\eta,k)$ but not $i\Delta(\eta,\vec{x};\eta,\vec{0})$.

If the first case is accepted (spatial Fourier transforms exist) then 
the TTS conditions completely fix the gauge and the on-shell field 
redefinition of Higuchi, Marolf and Morrison \cite{Higuchi:2011vw} is 
not allowed. In that case the linearized Weyl-Weyl correlator can 
indeed be used to derive a unique result (\ref{power}) for the tensor 
power spectrum, but it also gives a unique result 
(\ref{TTS1}-\ref{TTS2}) for the propagator, with the de Sitter breaking 
secular growth.

On the other hand, if the validity of spatial Fourier transforms is
denied, then the spatial Fourier transform of the linearized Weyl-Weyl 
correlator does not give a unique result for the tensor power spectrum. 
It must be supplemented by some boundary condition at spatial infinity
which defines how to invert $\nabla^2 \rightarrow -k^2$. In that case 
the linearized Weyl-Weyl correlator does not contain all physical 
information about free gravitons, and there is no significance to the 
fact that this correlator fails to show the de Sitter breaking secular 
dependence of the TTS gauge propagator (\ref{TTS1}-\ref{TTS2}).

If spatial Fourier transforms do not exist one must also accept that 
the graviton field operator possesses degrees of freedom in addition to 
the creation and annihilation operators $\alpha(\vec{k})$ and 
$\alpha^{\dagger}(\vec{k})$. What Higuchi, Marolf and Morrison did
instead is to retain only these degrees of freedom and reshuffle how
$h_{ij}(t,\vec{x})$ depends upon them \cite{Higuchi:2011vw}. That 
corresponds to a noncanonical quantization, not a fixing of some 
residual gauge freedom \cite{Miao:2011ng}.

Finally, it is important to note that there should be no local test
to distinguish between the manifold $R^3 \times R$, on which the
existence of spatial Fourier transforms might be an issue, and the
manifold $T^3 \times R$, on which spatial Fourier transforms certainly
exist. On the manifold $T^3 \times R$ the TTS gauge propagator becomes
a mode sum to which the continuum result (\ref{TTS1}-\ref{TTS2}) is
an excellent approximation for spatial coordinate separations which 
are small compared to the $T^3$ radii. As pointed out above, this 
propagator will certainly exhibit the de Sitter breaking secular 
growth which is at issue.

\subsection{Asymptotic symmetries are not fixed in flat space}

The debate over de Sitter breaking sometimes engenders a sense of
{\it d\'ej\`a vu}. The issue of spatially growing coordinate 
redefinitions has already come up --- quite a long time ago 
\cite{Bondi:1962px,Sachs:1962wk} --- in the context of graviton 
scattering amplitudes on flat space background. These BMS 
transformations act nontrivially on scattering amplitudes and are not
considered to be gauge transformations \cite{Strominger:2014pwa,
Strominger:2013jfa,He:2014laa}. In particular, no one employs them to
alter the infrared behavior of the graviton propagator, which cannot
be changed because it has physical consequences \cite{Weinberg:1965nx}.
Instead the action of a BMS transformation on a classical configuration
corresponds to the physical differences which prevail after a 
gravitational wave has passed \cite{Strominger:2014pwa}. {\it Note 
that the linearized Weyl curvature vanishes before and after the 
passage of such a wave.}

It is difficult to discern any difference of principle between the 
BMS transformations of graviton scattering on flat background and the
analogous spatially growing symmetries of cosmology worked out by
Hinterbichler, Hui and Khoury \cite{Hinterbichler:2013dpa}, the first 
of which was exploited by Higuchi, Marolf and Morrison 
\cite{Higuchi:2011vw}. It would seem to follow that these transformations 
should not be regarded as gauge symmetries, nor should they be employed 
to alter the infrared properties of the graviton propagator. The same
conclusion follows from working on the spatial manifold $T^3 \times R$, 
which has no extra symmetries.

\subsection{Things couple to the metric, not the Weyl tensor}

Another debate physics has already seen is whether or not undifferentiated
gauge fields can mediate physical effects when the field strength vanishes. 
It was obvious from the first charged particle wave equations of the 1920's 
that matter fields do not couple to the electromagnetic field strength 
tensor $F_{\mu\nu}$ but rather to the undifferentiated vector potential 
$A_{\mu}$. Hence there must sometimes be electromagnetic effects in regions 
throughout which the field strength vanishes. But physicists are as prone 
to feel prejudice as any other humans, and this obvious conclusion was 
denied for decades. Then came the work of Ehrenberg and Siday 
\cite{Ehrenberg:1948}, followed by that of Aharonov and Bohm 
\cite{Aharonov:1959fk}. Everyone knows how the experiment turned out 
\cite{Chambers:1960}.

The analogy to gravity seems obvious. The fact that the secular growth 
of the TTS propagator (\ref{TTS1}-\ref{TTS2}) takes the form of a 
linearized gauge transformation (in a completely gauge-fixed result) 
explains why it drops out of the linearized Weyl-Weyl correlator. 
However, that does not mean this time dependence is unphysical, any more
than the vanishing electromagnetic field strength implies that the
Aharonov-Bohm potential is unphysical. Matter, and gravity itself, 
couples to the metric, not to the linearized Weyl tensor, so there must 
be circumstances under which the metric can communicate physical 
effects even when the linearized Weyl tensor vanishes. One of these may
already have been found in a recent one graviton loop computation of 
the vacuum polarization on de Sitter background (cf. section 3.5 of) 
\cite{Glavan:2015ura}.

\section{Conclusion}

The decades over which varying opinions have been expressed about de 
Sitter breaking in the graviton propagator 
\cite{Antoniadis:1986sb,Allen:1986tt,Tsamis:1992xa,Kleppe:1993fz,
Hawking:2000ee,Higuchi:2001uv,Kouris:2001hz,Higuchi:2002sc,
Woodard:2004ut,Higuchi:2008fu,Faizal:2008ns,Marolf:2008hg,Miao:2009hb,
Miao:2010vs,Miao:2011fc,Faizal:2011iv,Higuchi:2011vw,Miao:2011ng,
Kahya:2011sy,Mora:2012kr,Higuchi:2012vy,Mora:2012zh,Mora:2012zi,
Morrison:2013rqa,Miao:2013isa,Bernar:2014lna,Frob:2014fqa} might make
the debate seem interminable. However, real progress has been made and
a definitive consensus has been reached on a number of essential 
issues:
\begin{itemize}
\item{All now agree on the validity of the Feynman rules used for all 
existing graviton loop computations on de Sitter \cite{Tsamis:1996qq,
Tsamis:1996qm,Tsamis:1996qk,Tsamis:2005je,Miao:2005am,Miao:2006gj,
Kahya:2007bc,Kahya:2007cm,Kitamoto:2012ep,Kitamoto:2012vj,Kitamoto:2012zp,
Leonard:2013xsa,Kitamoto:2013rea,Glavan:2013jca,Kitamoto:2014gva,
Wang:2014tza,Glavan:2015ura,Chu:2015ila};}
\item{All now agree that the open and closed coordinate mode sums show 
the same secular growth;}
\item{All now agree that it is valid to regard de Sitter as a special
case of inflationary cosmology in open coordinates;}
\item{The crucial importance of $\Delta^2_h(\eta,k)$ has been 
recognized; and}
\item{The significance of large scale transformations has been 
recognized.}
\end{itemize}
This has been achieved by members of the different communities 
thoughtfully considering each other's arguments. Fr\"ob's study makes 
a fine addition, and I have tried to  reply in the same spirit. 

Through Morrison's work we can now quantify the rather small 
differences between the de Sitter invariant and breaking propagators 
\cite{Morrison:2013rqa}. These differences survive in the one graviton 
loop correction to the vacuum polarization (cf. section 3.5 of) 
\cite{Glavan:2015ura}. What remains is to achieve a consensus on 
observables so that the physical significance of this result can be 
determined.

\centerline{\bf Acknowledgements}

I am grateful for conversations and correspondence on this subject with 
A. Fr\"ob, A Higuchi, S. P. Miao and G. Pimentel. This work was partially 
supported by NSF grants PHY-1205591 and PHY-1506513, and by the Institute 
for Fundamental Theory at the University of Florida.

\end{document}